# Field- and current-driven magnetic domain-wall inverter and diode

Zhaochu Luo[1,2*], Stefan Schären[1,2], Aleš Hrabec[1,2,3],

Trong Phuong Dao[1,2,3], Giacomo Sala[3], Simone Finizio[2], Junxiao Feng[3],

Sina Mayr[1,2], Jörg Raabe[2], Pietro Gambardella[3*], Laura J. Heyderman[1,2*]

[1]Laboratory for Mesoscopic Systems, Department of Materials, ETH Zurich, 8093 Zurich, Switzerland.
[2]Paul Scherrer Institut, 5232 Villigen PSI, Switzerland.
[3]Laboratory for Magnetism and Interface Physics, Department of Materials, ETH Zurich, 8093 Zurich, Switzerland.

**ASTRACT**

We investigate the inversion process of magnetic domain walls (DWs) propagating through synthetic noncollinear magnetic textures, whereby an up/down DW can be transformed into a down/up DW and vice versa. We exploit the lateral coupling between out-of-plane and in-plane magnetic regions induced by the interfacial Dzyaloshinskii–Moriya interaction in Pt/Co/AlOx trilayers to realize both field-driven and current-driven magnetic DW inverters. The inverters consist of narrow in-plane magnetic regions embedded in out-of-plane DW racetracks. Magnetic imaging and micromagnetic simulations provide insight into the DW inversion mechanism, showing that DW inversion proceeds by annihilation of the incoming domain on one side of the in-plane region and nucleation of a reverse domain on the opposite side. By changing the shape of the in-plane magnetic region, we show that the DW inversion efficiency can be tuned by adjusting the ratio between the chiral coupling energy at the inverter boundary and the energy cost of nucleating a reverse domain. Finally, we realize an asymmetric DW inverter that has nonreciprocal inversion properties and demonstrate that such a device can operate as a DW diode. Our results provide input for the versatile manipulation of DWs in magnetic racetracks and the design of efficient DW devices for nonvolatile magnetic logic schemes.

**Main text:**

I. <u>Introduction</u>

Magnetic domain-walls (DWs) constitute a physical medium to encode digital information as an alternative to the electron charge in electronic circuits. DWs can be displaced either by applying a magnetic field or by injecting an electric current. Following early attempts to realize memories based on DW motion in thin films using Oersted fields [1,2], in 2008 Parkin *et al.* proposed a DW racetrack memory employing current-induced displacement of DWs along nano-sized magnetic wires [3,4]. In the last decade, the racetrack concept has been significantly developed to achieve higher energy efficiency and faster DW motion by exploiting improved material systems [5-8] and different physical phenomena that contribute to the DW motion [9-15]. In particular, spin-orbit torques (SOTs) arising from the spin Hall and Rashba-Edelstein effect have proved to be an effective way to drive the DW motion with an electric current [16].

DW-based logic architectures capable of performing Boolean logic operations have been developed in parallel to DW memory devices [1,17-21]. In 2005, Allwood *et al.* demonstrated elementary logic operations using DWs in magnetic circuits with in-plane magnetization by applying a rotating magnetic field [18,19]. This work exemplified key advantages of DW logic in terms of fan-out ability, cascading, low energy consumption, and built-in non-volatile memory. However, in order to realize efficient and scalable DW logic circuits, it is desirable to (i) migrate this concept to a perpendicular magnetic system that would enable a higher DW density and faster DW motion [15], and (ii) achieve logic operations using an electric current rather than an external magnetic field, which would allow for individual addressing of the gates and minimal energy consumption.

In our recent work, we have targeted these two challenges by exploiting the chiral coupling between adjacent magnets with competing magnetic anisotropy and interfacial Dzyaloshinskii–Moriya interaction (DMI) [22,23], and have demonstrated a current-driven DW inverter as well as reconfigurable NAND and XOR gates in perpendicular magnetic racetracks, which can be cascaded to perform different logic operations [24]. In this paper, we provide a deeper insight into the mechanism of the DW inversion and demonstrate how to construct a magnetic field-and current-driven DW inverter in a system with perpendicular anisotropy. We then tailor the symmetry of the inverter to introduce non-reciprocity in the DW inversion process and show that such a device can operate as a DW diode. Our findings exemplify the versatility of DW logic elements based on chirally coupled nanomagnets for both field- and current-driven operation.

II. Experimental methods

The DW devices were fabricated from a Pt (6 nm)/Co (1.6 nm)/Al (2 nm) trilayer, which has a large DMI at the Pt/Co interface [25,26] as well as tunable magnetic anisotropy that depends on the degree of oxidation of the Co/Al interface [27-29]. Specifically, the magnetization direction can be tuned from in-plane (IP) to out-of-plane (OOP) by oxidizing Al into $AlO_x$.

The magnetic trilayers were deposited on a 200 nm-thick $SiN_x$ layer on a silicon substrate using dc magnetron sputtering with a base pressure $<2\times10^{-8}$ Torr and a deposition Ar pressure of 3 mTorr. The as-grown films have an IP magnetic anisotropy and the patterning of these films into DW devices was carried out by electron-beam lithography. First, the continuous films were milled into strips with Ar ions through a negative resist (ma-N2401) mask. In these magnetic strips, the upper Co/Al bilayer was milled through a high-resolution positive resist (PMMA) mask to create the DW racetracks. The milling process is stopped before reaching the Pt layer, which is used to provide a uniform electric conducting channel. In order to define the regions with IP and OOP magnetic anisotropies in the racetracks, a second PMMA mask was patterned by electron-beam lithography on top of the Al layer. Using a low-power (30 W) oxygen plasma at an oxygen pressure of 10 mTorr, the unprotected Al layer was oxidized to induce perpendicular magnetic anisotropy in the required regions in the Co layer as a result of the Co/AlOx interface. Finally, electrodes of Cr (5 nm)/Au (50 nm) were fabricated using electron-beam lithography combining electron-beam evaporation with a lift-off process.

The DW motion in the magnetic racetracks was measured using polar magneto-optical Kerr effect (MOKE) microscopy. First, a background image was captured after saturating the OOP magnetization using an OOP magnetic field of 1 kOe. The background image was then subtracted from the subsequent images to achieve differential images with magnetic contrast. Thus, the dark and bright contrast in the differential images correspond to down (⊗) and up (⊙) magnetization. The DW motion was driven either by magnetic field or by the current pulses of variable amplitude and width. Furthermore, high-resolution imaging of the DW inverter was performed using scanning transmission x-ray microscopy (STXM) at the PolLux beamline of the Swiss Light Source. The magnetization state was probed by exploiting the x-ray magnetic circular dichroism (XMCD) at the Co $L_3$ absorption edge at normal incidence. The devices measured using STXM were fabricated on x-ray transparent $SiN_x$ membranes.

III. Principle of chiral coupling

The DMI arises in magnetic systems that lack a center of inversion symmetry and exhibit strong spin–orbit coupling. This interaction favors the orthogonal alignment of adjacent magnetic moments with a fixed chirality, as expressed by the Hamiltonian $H_{DMI} = -\boldsymbol{D} \cdot (\boldsymbol{m}_1 \times \boldsymbol{m}_2)$, where $\boldsymbol{D}$ and $\boldsymbol{m}_i$ ($i$ = 1 or 2) represent the DMI vector and two neighboring magnetic moments, respectively [30,31]. In competition with the exchange interaction, magnetic anisotropy and dipolar interaction, the DMI favors the formation of noncollinear spin textures such as chiral Néel domain walls [10-12,32-35], spin helices, cycloids, and skyrmions [36-39]. Here we exploit the lateral chiral coupling promoted by the DMI in combination with patterning of the magnetic anisotropy of a Pt/Co/Al trilayer in order to achieve control over the preferred magnetic configurations of consecutive regions of a racetrack with IP and OOP magnetization. As a result of the DMI arising from the Pt/Co interface, the left-handed configurations down-right (⊗→) and up-left (⊙←) are energetically favored over the right-handed configurations ⊗← and ⊙→ in adjacent OOP-IP regions. By patterning the magnetic anisotropy with high-resolution lithography, we are able to create chirally coupled nanomagnetic systems with extended OOP-IP structures that can be used for fabricating artificial spin ices [40], lateral synthetic antiferromagnets, synthetic skyrmions, and field-free memory elements [22] as well as current-driven DW injectors [33] and DW logic circuits [23,24].

We consider here an OOP racetrack with an embedded IP region, where chiral coupling induces antiparallel alignment of the OOP magnetization on the two sides of the IP region while the IP magnetization satisfies the left-handed chirality rule. This structure can be considered as a "pinned" artificial Néel-type DW, with preferred OOP-IP-OOP magnetic orientations given by ⊗→⊙ or ⊙←⊗ (Fig. 1a). In the following, we show that by applying a rotating magnetic field, we can realize the inversion of a DW propagating through such an OOP-IP-OOP structure (Fig. 1b).

To demonstrate the operation of the perpendicular magnetic DW inverter with a magnetic field, we fabricated a 2 μm-wide ring-shaped magnetic racetrack, previously proposed for memory applications [3, 41], with an embedded 4 μm-long IP region (Fig. 2a). The sequence of MOKE images (b) to (i) in Fig. 2 shows the DW inversion on applying 1 s-long magnetic field pulses of 300 Oe in the *x-z* plane. The initial magnetization configuration shown in the MOKE image (b) is indicated in the schematic in Fig. 2a. On applying $+H_z$ pulses, the ⊙ domain starts to expand, leading to the anti-clockwise propagation of the ⊙|⊗ DW (indicated in red) along the racetrack (Fig. 2c). Once it reaches the left side of the IP region, the ⊙|⊗ DW annihilates (Fig. 2d). Then a $-H_x$ magnetic field pulse is applied to switch the IP magnetization from → to ←. Due to the chiral coupling, the OOP magnetization on both sides of the IP region experiences an effective DMI field given by

$$\vec{H}_{DMI} = \frac{2D}{\mu_0 M_S}\left(-\frac{\partial m_z}{\partial x}, 0, \frac{\partial m_x}{\partial x}\right), \quad (1)$$

where $\mu_0$ is the vacuum permeability, $M_S$ is the saturation magnetization, and $m_z$ and $m_x$ are the *z* and *x* components of the magnetization, respectively. Hence, $H_{DMI}$ contains a *z*-component pointing to ⊙ and ⊗ for the OOP magnetization on the left and right side of the ← IP region, respectively. The subsequent magnetic field pulses along $-H_z$ facilitate the nucleation of the ⊗ domain to the right of the IP region and create the inverted ⊗|⊙ DW that will propagate anti-clockwise along the racetrack (Fig. 2e and 2f). Once the ⊗|⊙ DW travels around the entire racetrack and annihilates to the left of the IP region, a $+H_x$ magnetic field pulse is applied to switch the IP magnetization from ← to → (Fig. 2g). In this case, $H_{DMI}$ contains a *z*-component pointing towards ⊗ and ⊙ for the OOP magnetization on the left and right side of the IP region, respectively. The subsequent $+H_z$ magnetic field pulses facilitate nucleation of the ⊙ domain to the right of the IP region and push the inverted ⊙|⊗ DW anti-clockwise along the magnetic racetrack (Fig. 2h and 2i). We thus reach a state that is equivalent to the initial state. Hence, by applying a magnetic field that rotates periodically in the *x-z* plane, the DW propagates anti-clockwise along the magnetic racetrack and is continuously inverted when propagating through the IP region, showing that the OOP-IP-OOP structure can serve as a DW inverter in systems with out-of-plane anisotropy operated by a magnetic field.

IV. Current-driven DW inversion

Controlling DW operations with an electric current rather than a magnetic field is more interesting for practical application in terms of addressability, scalability, and energy efficiency. It has been extensively shown that SOTs provide an efficient mechanism to displace the Néel-type DWs of both polarities using an electric current [9-12,16]. The effective magnetic field resulting from the damping-like SOT responsible for DW motion is given by:

$$\vec{H}_{SOT} = \frac{\hbar \theta_{SH} J}{2\mu_0 e M_S t} \vec{m} \times \vec{\sigma}, \tag{2}$$

where $\hbar$ is the reduced Planck constant, $\theta_{SH}$ is the effective spin Hall angle, $J$ is the electric current density, $e$ is electron charge, $t$ is the thickness of the magnetic layer and $\sigma$ is the direction of the spin polarization at the Pt/Co interface, respectively. As shown previously, the DMI in Pt/Co/AlO$_x$ promotes the left-handed Néel-type DW. Because the magnetization in the middle of the ⊙|⊗ (⊗|⊙) DW points ← (→), $H_{SOT}$ points to ⊙ (⊗), resulting in the expansion of the ⊙ (⊗) domain along the direction of the electric current (Fig. 3a and 3b).

We have shown in our previous work that when the IP region is sufficiently narrow, the incident DW driven by SOTs inverts when it reaches the IP region, so accomplishing the current-driven DW inversion operation [24]. We now analyze the SOT-driven DW inversion process in more detail as shown schematically in Fig. 3. When a ⊗|⊙ DW propagates from left to right in Fig. 3a and 3b, and approaches the IP region, the magnetization in the IP region experiences a dipolar field $H_{dip}$(IP) generated by the IP magnetization of the ⊗|⊙ DW that

points → (Fig. 3c). As $H_{SOT}$ compresses the incident DW against the IP region, the dipolar energy increases as well as the exchange energy due to the decreased domain size. This results in a compact, high-energy spin texture containing two closely spaced regions with head-to-head IP magnetization, as shown by the associated magnetic charges in Fig. 3c: one IP magnetization region is in the middle of the ⊗|⊙ DW with magnetization pointing →, and the other IP magnetization region is in the inverter (blue shaded region) and has ← pointing magnetization. When the SOT pushes the DW against the IP region, the dipolar field becomes strong enough to switch the narrow IP region from ← to → with the help of SOTs (Fig. 3d). Simultaneously, the ⊙ domain on the left side of the IP region annihilates. After the reversal of the IP magnetization, a reverse ⊙ domain is nucleated on the right side of the IP region as a consequence of $H_{DMI}$(OOP) pointing towards $+z$. The magnetization points ← in the middle of the resulting ⊙|⊗ DW and $H_{SOT}$(DW) points towards $+z$ so that this DW is then propagated onwards by the electric current. The inversion process also works for an incident ⊙|⊗ DW as well as for the opposite direction of the current. Therefore, such an OOP-IP-OOP structure serves as a current-driven DW inverter.

V. Symmetric DW inverter

In order to tailor the performance of the inverter, we tested different shapes of the IP region, namely straight regions orthogonal to the racetrack and V-shaped regions rotated by 90° with respect to the racetrack. In the following, we show that the V-shaped inverter performs better in terms of reliability of the inversion process. We also show that the DW inversion process is bi-directional, *i.e.* the DW can be inverted in both the forward and backward propagation directions. The operation of a V-shaped DW inverter with a 50 nm-wide IP region is shown in Fig. 4a. When the DW arrives at the IP region, it is annihilated while a reverse domain nucleates on the other side, creating a DW with inverted polarity [24]. We observe a similar inversion process for a DW propagating in the opposite direction when reversing the direction of the electric current.

The details of the bi-directional DW inversion mechanism can be directly imaged by using scanning transmission x-ray microscopy (STXM). As shown in Fig. 4b, when the DW reaches the IP region from the left side of the "V", the magnetization on both sides of the IP region points in the same direction, forming a DMI-unfavorable high-energy configuration. This configuration can only be unwound by nucleating a reverse domain on the right side of the IP region. At the right side of the apex of the "V", the OOP region is surrounded by the IP region, providing an artificial nucleation center for the reverse domain.

The probability of nucleating a reverse domain along the IP region depends on the energy gain/cost ratio $\lambda$ relative to the situation in which the two OOP regions on either side of the IP region have parallel magnetization. To model $\lambda$ in a simple way, we consider the creation of a fan-shaped domain of radius $R$ and cone angle $\theta$ along the IP region. The energy gain $E_{\text{gain}} = NR\sigma_{\text{DMI}}$ is resulted from satisfying the DMI-induced chiral coupling that occurs at the boundary of the IP region, and the energy cost $E_{\text{cost}} = \theta R \sigma_{\text{DW}}$ is associated to the DW energy in the OOP region, where $N$ is the number of IP boundaries surrounding the apex of the fan-shaped domain, $\sigma_{\text{DMI}}$ is the chiral coupling energy per unit length and $\sigma_{\text{DW}}$ is the DW energy per unit length. Because the chiral coupling is significantly stronger than the dipolar interaction in this system [22], the energy gain due to the dipolar interaction is neglected. Thus, the energy ratio $\lambda$ for nucleating a reverse domain is:

$$\lambda = \frac{E_{\text{gain}}}{E_{\text{cost}}} = \frac{NR\sigma_{\text{DMI}}}{\theta R \sigma_{\text{DW}}} \sim \frac{N}{\theta}. \tag{3}$$

As shown in the schematics in Fig. 4c, $N = 1$ for the nucleation position at the intersection of the "V" with the racetrack edge, and $N = 2$ for other nucleation positions along the "V". Referring to the schematics in Fig. 4c, we define $\alpha$ as the tilt angle of the IP region relative to the racetrack direction. We find that $\lambda \sim 1/\alpha$ when $N = 2$ and $\theta = 2\alpha$ at the right side of the apex of the V shape, whereas $\lambda \sim 1/(\pi-\alpha)$ when $N = 1$ and $\theta = \pi-\alpha$ at the intersection of the "V" with the racetrack edge and $\lambda \sim 2/\pi$ where $N = 2$ and $\theta = \pi$ at the nucleation positions along the arms of the "V". Because $\alpha < \pi/2$, the nucleation probability is higher at the apex of the "V", and the area on the right side of the apex of the V-shaped region serves as an artificial nucleation center for a positive electric current. When the direction of the electric current is reversed, the DW reaches the IP region from the right side of the "V", thus forming a DMI-unfavorable high-energy configuration. In this case, this configuration can be unwound by nucleating a reverse domain on the left side of the IP region. At the left side of the intersection of the "V" with the edge of the racetrack, we have $N = 1$ and $\theta = \alpha$, which gives $\lambda \sim 1/\alpha$. This is larger than $\lambda \sim 1/(\pi-\alpha)$ at the left side of the apex of the "V", and $\lambda \sim 2/\pi$ along the arms of the "V" (see the schematics in Fig. 4c). Hence, the intersections of the "V" with the edge of the racetrack serves as an artificial nucleation center for the reverse domain with a negative electric current.

We now investigate the effect of the angle $\alpha$ on the performance of the inverter by fabricating V-shaped DW inverters with various $\alpha$ and a constant width of the IP region. In particular, we determine the probability of nucleating a reverse domain starting from the initial inverter configuration of $\odot \leftarrow \otimes$ after applying fifty 50 ns-long pulses at constant current density $j = 4.2 \times 10^{11}$ A/m$^2$. The nucleation probability is determined for each angle by taking the average of 100 trials of DW inversion operations measured with MOKE microscopy. As shown in Fig. 5, the nucleation probability increases with decreasing $\alpha$, which corresponds to a higher energy ratio $\lambda$, and reaches 100% when $\alpha = 0.1\pi$. This result implies that for $\alpha < \alpha_C = 0.1\pi$ the energy gain for nucleating a reverse domain is larger than the energy cost, i.e. $\lambda > 1$, leading to a deterministic nucleation process. This critical angle can be deduced from Eq. 3: $\alpha_C = \sigma_{DMI} / \sigma_{DW}$. By taking the experimental parameters [24], we estimate that the DW energy density is

$$\sigma_{DW} = \left(4\sqrt{AK_{eff}} - \pi D\right) t \sim 6.3 \times 10^{-12} \text{ J/m},$$

and the chiral coupling energy density, i.e. the energy difference between the transient parallel ($\odot \rightarrow \odot$ or $\otimes \leftarrow \otimes$) and antiparallel ($\odot \leftarrow \otimes$ or $\otimes \rightarrow \odot$) configuration [22] is

$\sigma_{DMI} = \pi |D| t \sim 4.5 \times 10^{-12}$ J/m, where $A$=16 pJ/m, $D$=0.9 mJ/m$^2$, $t$=1.6 nm, $K_{eff}$=180 kJ/m$^3$. So, the energy gain/cost model gives the critical angle $\alpha_C = 0.71$, which is about a factor two larger than the experimental value of 0.31. This discrepancy can be ascribed to the uncertainty in some of the micromagnetic parameters. In addition, in this model, the process of formation of the transient configuration with parallel OOP magnetization (see the schematic in Fig. 3c) is not considered. In particular, the incident DW needs to overcome an energy barrier to annihilate and form the parallel configuration. This energy barrier depends on the spin texture surrounding the IP region, i.e. the geometry of the IP region. The small tilt angle of the magnetization in the IP region will give smaller repulsive dipolar interaction with the incident DW and facilitate the annihilation process. Moreover, we observe a slight decrease of the nucleation probability for a positive electric current compared with a negative electric current, which we ascribe to the difference of magnetization dynamics in the two cases that is not considered in our model (see Section VII).

VI. Asymmetric DW inverter

We now show that it is possible to introduce non-reciprocity in the DW inversion operation by modifying the shape of the DW inverter. As shown in Fig. 6a, we increase the width of the racetrack on the left side of the IP region to quench the probability for domain nucleation for a negative electric current. In this case, the OOP region at the intersection of the "V" with the racetrack has a smaller nucleation probability than the other positions along the IP region because $\lambda \sim 1/(\pi/2+\alpha) < 2/\pi$. Thus, it becomes difficult to nucleate a reverse domain on the left side of the IP region. We have experimentally demonstrated this point by propagating DWs from the right towards the left side of the inverter using a negative electric current (Fig. 6b). When the DWs reach the V-shaped IP region from the right, they remain pinned and cannot pass through the inverter. In contrast, DWs incident from the left can be transmitted towards the right side with a positive current because the nucleation center on the inner side of the apex of the "V" is unaffected. This behavior is observed for both $\odot|\otimes$ and $\otimes|\odot$ DWs (Fig. 6b). Therefore, the nucleation of a reverse domain is highly asymmetric for positive and negative electric currents but independent of the DW polarity. We note also that change in current density required when reversing the current, which is a result of the asymmetric structure of the Co layer, is small ($\approx 6\%$), because most of the current flows through the Pt channel. Thus, the difference of the SOTs acting on the DW in the wide and narrow side of the racetrack is minor. Moreover, on measuring the DW displacement as a function of current density, we find that the DW propagates through the asymmetric inverter with a similar velocity to that of the DW passing through the symmetric inverter in the forward direction, whereas it is completely hindered in the backward propagation direction (Fig. 6c).

VII. Micromagnetic simulations

To further understand the mechanism of the DW inversion in both symmetric and asymmetric DW inverters, we performed micromagnetic simulation using MuMax$^3$ [42]. The simulation contains 2048×1024×1 cells with a 2×2×1.6 nm$^3$ discretization using the following magnetic parameters: saturation magnetization $M_S$ = 0.9 MA/m, effective OOP anisotropy field $H_{eff}$ = 200 mT, exchange constant $A$ = 16 pJ/m, effective spin Hall angle Pt $\theta_{sh}$ = 0.1, and interfacial DMI constant $D$ = -1.5 mJ/m$^2$. The spin current is uniformly applied across the inverter structure. The width of the racetrack in the symmetric DW inverter is 800 nm, and the width of the left and right racetracks in the asymmetric DW inverter is 1500 nm and 800 nm, respectively. The width of the IP region is 30 nm and $\alpha$ = 20°.

The simulations of the current-driven DW inversion in both symmetric and asymmetric DW inverters are shown in Fig. 7. The current density-velocity curves exhibit three regimes of different DW behaviors (Fig. 7a). In regime I at low current density ($j$ < 1.0×10$^{12}$ A/m$^2$), the DW is pinned just before the IP region both in the symmetric and asymmetric DW inverters irrespective of the current direction (see snapshots I in Fig. 7b). Pinning is attributed to the energy barrier that must be overcome to switch the chiral magnetic texture surrounding the inverter region. For example, as shown in Fig. 7b, if the initial magnetic configuration of the inverter is ⊗→⊙, as favored by the DMI, the inverter has to switch to the other DMI-favored configuration ⊙←⊗ in order to transfer the incident DW from the left to the right side.

In regime II, *i.e.* at moderate current density (1.0×10$^{12}$ < $j$ < 2.0×10$^{12}$ A/m$^2$), the SOT is strong enough to push the DW across the energy barrier and nucleate a reverse domain on the other side of the IP region. In this regime, the speed of the DW through the inverter increases linearly with the current density. The reverse domain in the symmetric DW inverter nucleates at the apex of the "V" for a positive electric current and at intersection of the "V" with the edge of the racetrack for a negative electric current, in agreement with the STXM measurements and the arguments presented in Sect. IV. In the simulation of the backward propagating DW across the symmetric DW inverter, we find that the nucleation process differs at the upper and bottom intersections of the "V" with the edge of the racetrack (see snapshots II in Fig. 7b), which is due to the DW tilt caused by the combined effect of SOTs and DMI [43,44]. Moreover, the threshold current density for the DW inversion in the forward propagation direction is slightly lower than that for the backward propagation direction, implying that, in spite of the same chiral coupling strength, the nucleation at the apex of the "V" is more favorable than that at the edges of the racetrack. In the simulation of the asymmetric DW inverter, the behavior in the forward propagation direction is the same as that of the symmetric DW inverter and their DW velocities almost coincide. In the backward propagation direction, there is no nucleation at the edges and the DW is fully blocked in front of the IP region. This non-reciprocal behavior is in very good agreement with the experimental results reported in Fig. 6.

Finally, regime III at high current density ($j > 2.0\times10^{12}$ A/m$^2$) corresponds to the formation of complex magnetic textures around the IP region (see snapshots III in Fig. 7b) and to the saturation of the DW velocity. With even higher current density ($j > 2.6\times10^{12}$ A/m$^2$), magnetic domains in the form of bubbles are continuously injected from the apex of the "V", leading to the breakdown of the DW inverter in the backward propagation direction (indicated by crosses in Fig. 7a).

Note that the quantitative discrepancies between the simulations and experimental results (Fig. 6c and Fig. 7a) arise from the simplifications made in the micromagnetic model. For example, we ignore the DW pinning, which prevents DW motion at moderate current density. Pinning is known to be relatively high in Pt/Co films [45]. In addition, the simulations are performed at 0 K so that any thermally induced processes are neglected.

## VIII. DW diode

In electronic circuits, the diode is a key non-reciprocal element that rectifies the electron flow, *i.e.* it allows for the electric current to pass in one direction (forward) while blocking it in the opposite direction (backward). This rectification property is commonly used to convert AC signals into DC signals in analog circuits. Previous work has shown that a similar rectification concept can be realized in field-driven magnetic DW circuits, but a current-driven DW diode has not been demonstrated so far [46-48]. Here we realize such a DW diode based on the non-reciprocal DW inverter described in Sect. VI.

The DW velocity-current density curves reported in Fig. 6c and 7a for the asymmetric DW inverter mimic the *I/V* characteristic of a diode. However, the DW propagating in the forward direction is inverted with respect to the initial DW. This effect has no counterpart in an electronic circuit. A correctly functioning DW diode can therefore be obtained by cascading one asymmetric DW inverter and one symmetric DW inverter. As shown in Fig. 8a, due to the chiral coupling, the magnetization on the left side and right side of the two cascaded DW inverters is the same in the equilibrium state. With a positive electric current, the DW that transfers through the asymmetric DW inverter with inverted polarity is inverted back at the symmetric inverter, so that the final DW propagates in the racetrack while maintaining its polarity. In contrast, with a negative electric current the DW can pass through the symmetric DW inverter with inverted polarity but it is blocked by the asymmetric DW inverter, as shown in Fig. 8b. This non-reciprocal operation is equivalent to a DW diode. The current-driven DW diode is a component that extends the operation of DW logic circuits to the AC signal regime.

IX. <u>Conclusions</u>

We have shown that the lateral chiral coupling induced by the interfacial DMI can be used to fabricate field-driven and current-driven DW inverters. The detailed mechanism of current-driven DW inversion has been observed using MOKE and high-resolution STXM imaging, with the DW inversion process proceeding by the annihilation of an incoming DW and nucleation of a DW of reverse polarity on the opposite side of the inverter IP region. The DW inversion is most effective for narrow V-shaped IP regions with a cone angle of about $0.1\pi$ due to the higher energy gain associated with DMI and lower energy cost of creating a DW in narrow constrictions. Depending on the direction of the current, the reverse domains nucleate either at the apex or at the edges of the V-shaped IP region with almost equal probability. By designing asymmetric DW inverters, we show that we can induce unidirectional propagation of DWs, so completely blocking the transmission of inverted DWs in one direction while allowing them to pass in the opposite direction. By combining an asymmetric DW inverter and a symmetric inverter, we have realized the analogue of a DW diode, which selectively allows for the propagation of DWs along one direction while maintaining their polarity. This device extends the capability of current-driven DW logic circuits [24] by enabling the rectification of DW propagation. Additionally, by tuning the asymmetry of the DW inverter, DW diodes can be used to clamp the propagation of DWs to a specific current level. As shown in the DW velocity curves in Fig. 7a, DWs can only propagate through the DW inverter when the current is positive and larger than a threshold value. In the same way, for the DW diode consisting of a symmetric and asymmetric DW inverter, DWs are only allowed to propagate at a current level larger than the threshold current. These properties enable the operation of DW logic circuits with AC currents as well as the implementation of reverse current protection schemes in magnetic racetracks. Finally, in addition to the rectification of DW propagation, DW diodes can be potentially used to fabricate simple OR and AND logic gates. These unique possibilities for DW manipulations allow for greater flexibility in the design of DW-based circuits.

The data that support this study are available via the Zenodo repository [49].

**Acknowledgements**

We thank A. Weber and V. Guzenko for technical support with sample fabrication and measurement. This work was supported by the Swiss National Science Foundation through grant No. 200020-172775. A.H. was funded by the European Union's Horizon 2020 research and innovation programme under the Marie Skłodowska-Curie grant agreement number 794207 (ASIQS). S.M. acknowledges funding from the Swiss National Science Foundation under Grant Agreement No. 172517. Part of this work was performed at the PolLux (X07DA) endstation of the Swiss Light Source, Paul Scherrer Institut, Villigen, Switzerland. The PolLux endstation was financed by the German Bundesministerium für Bildung und Forschung under Grant agreement No. 05KS4WE1/6 and 05KS7WE1.

**Figures and figure captions**

**Figure 1**

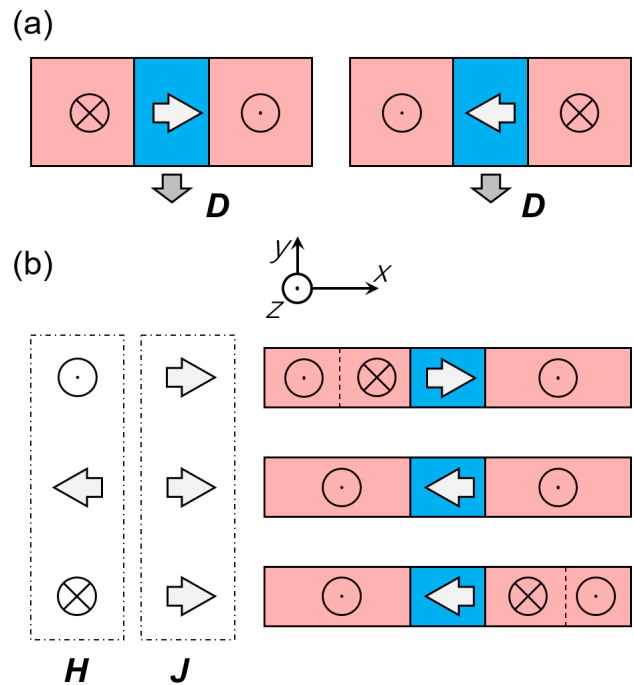

**Fig. 1.** Principle of chiral coupling and DW inversion. (a) Top view schematic of the chirally-coupled OOP-IP-OOP structure in an OOP racetrack showing the two magnetic configurations favored by the DMI, namely ⊗→⊙ (left) and ⊙←⊗ (right). The red and blue shaded areas correspond to the regions with OOP and IP magnetic anisotropy, respectively. The resultant DMI vector ***D*** for the Pt/Co/AlOx interface is indicated. (b) Schematics showing the DW inversion process on application of a magnetic field ***H*** rotating in the *x-z* plane or a positive electric current ***J*** along the *x*-axis.

**Figure 2**

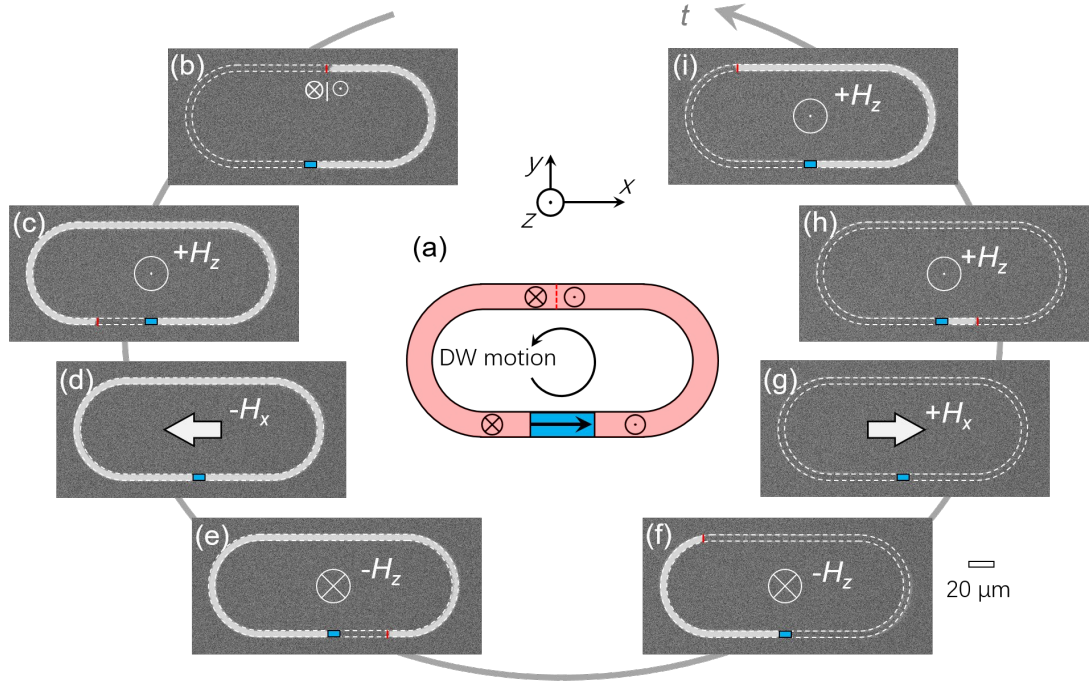

**Fig. 2.** DW inversion driven by a magnetic field. (a) Schematic of the field-driven DW inverter. The red and blue shaded areas correspond to the regions with OOP and IP magnetic anisotropy, respectively. (b-i) Differential MOKE images showing the operation of a field-driven DW inverter. The initial state in (b) corresponds to the schematic shown in (a). Images (c) to (i) show the magnetic configuration after applying a sequence of magnetic fields $+H_z$, $-H_x$, $-H_z$, $+H_x$, $+H_z$, … with $|H_{z,x}|$ = 300 Oe. The bright and dark contrast in the racetrack corresponds to $\odot$ and $\otimes$ magnetization, respectively. The red lines depict the positions of the DW.

**Figure 3**

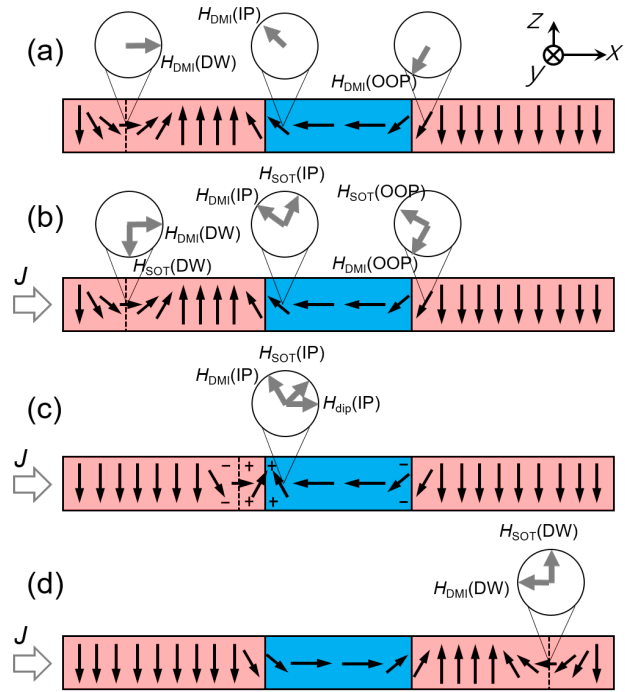

**Fig. 3.** Mechanism of current-driven DW inversion in an OOP-IP-OOP structure. (a-d) Side view schematics of the DW transfer through the inverter. The directions of the effective fields $H_{DMI}$, $H_{SOT}$, and $H_{dip}$ are indicated by arrows in the different regions. The magnetic charges associated with the compact high-energy spin texture are shown in (c).

**Figure 4**

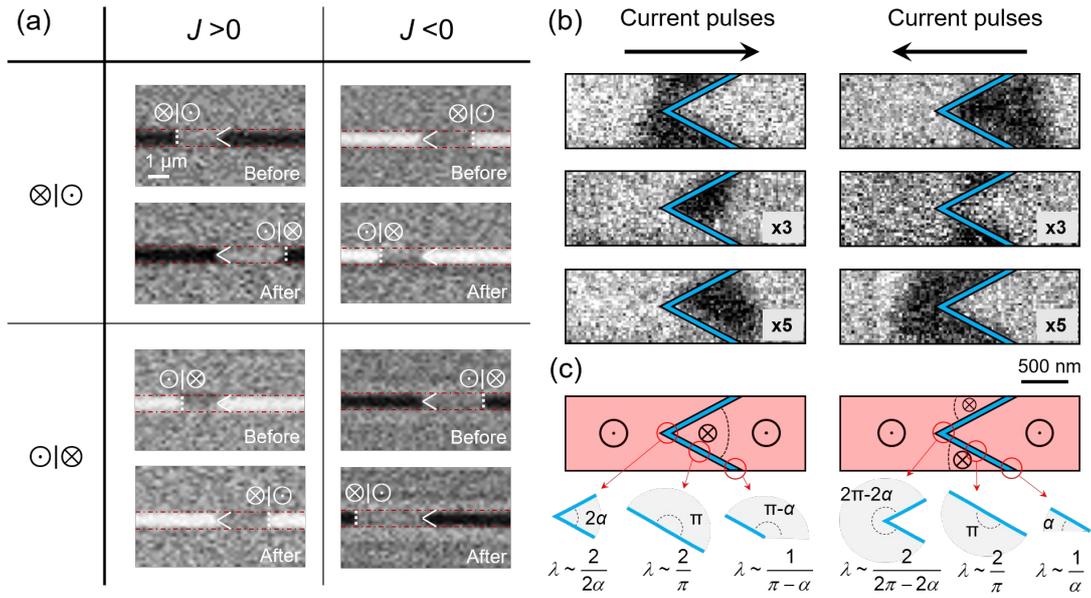

**Fig. 4.** Bi-directional operation of the current-driven DW inverter with a V-shaped IP region. (a) Differential MOKE images of the DW inverter before and after applying 30 electric current pulses. The bright and dark contrast in the MOKE images corresponds to ⊙ and ⊗ magnetization, respectively. (b) Sequence of XMCD images acquired during the DW inversion measured by STXM. The bright and dark contrast in the XMCD images corresponds to ⊙ and ⊗ magnetization, respectively. The current density and duration of the current pulses in (a) are $4.2 \times 10^{11}$ A/m$^2$ and 50 ns, whereas in (b) they are $1.1 \times 10^{12}$ A/m$^2$ and 1 ns. The numbers in the boxes represent the number of current pulses applied before the acquisition of each image. (c) Schematics showing a top view of the inverter with the fan angle of the inverted domain at different positions along the "V" for a positive (left schematic) and negative (right schematic) electric current. The energy ratio defined by equation (3) for nucleating a reverse domain at each position is indicated.

**Figure 5**

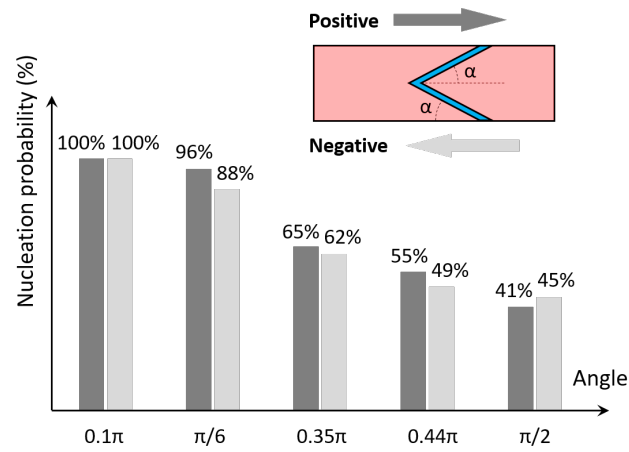

**Fig. 5.** Probability of nucleation of a reverse domain for different angles $\alpha$ (the definition of the angle $\alpha$ of the tilted IP region is indicated in the inset). The reverse domain nucleation is measured after applying 50 current pulses with current density $j = 4.2\times10^{11}$ A/m$^2$ with positive (dark gray) and negative (light gray) direction.

**Figure 6**

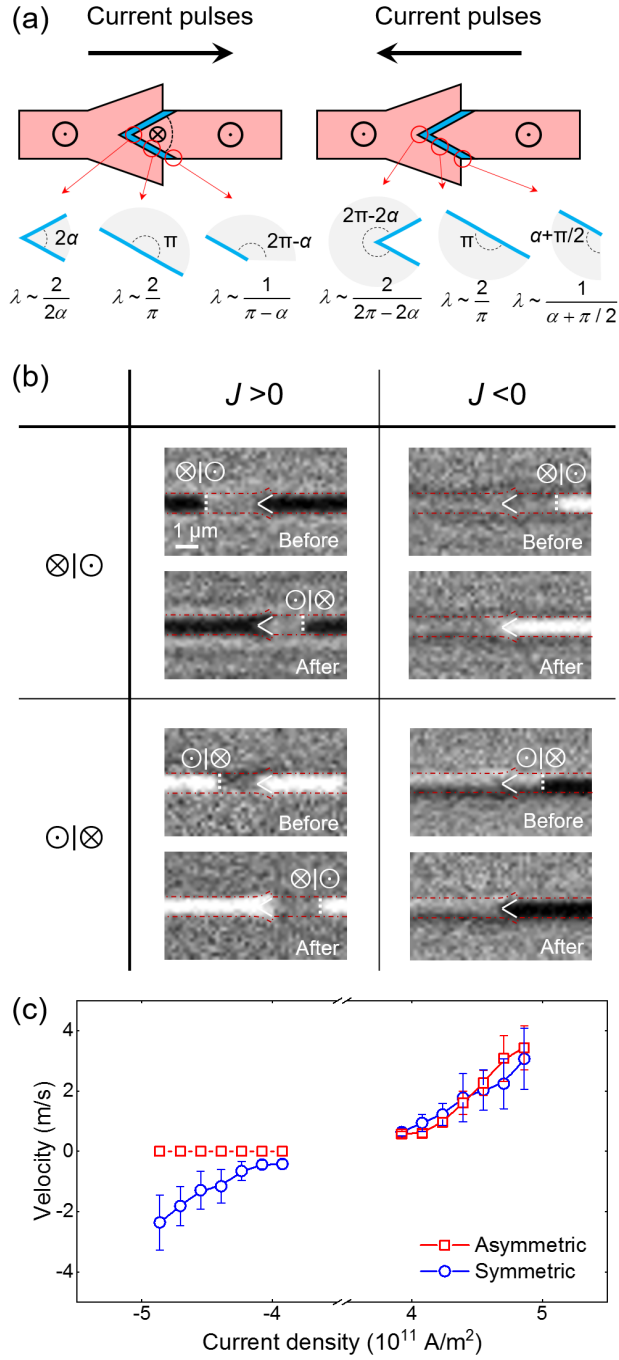

**Fig. 6.** Non-reciprocal operation of the asymmetric DW inverter. (a) Schematics showing a top view of the asymmetric DW inverter with the fan angle of the inverted domain at different positions along the "V" for a positive (left schematic) and negative (right schematic) electric current. The energy ratio defined by equation (3) for nucleating a reverse domain at each position is indicated. (b) Differential MOKE images of the asymmetric DW inverter before and after applying 30 electric current pulses. The bright and dark contrast in the MOKE images correspond to $\odot$ and $\otimes$ magnetization, respectively. The current density and the duration of the current pulses are $4.2\times10^{11}$ A/m$^2$ and 50 ns, respectively. (c) Velocity of DWs transferring through the asymmetric and symmetric DW inverters as a function of current density. Error bars represent the standard deviation of the DW velocity measured in 5 different devices.

**Figure 7**

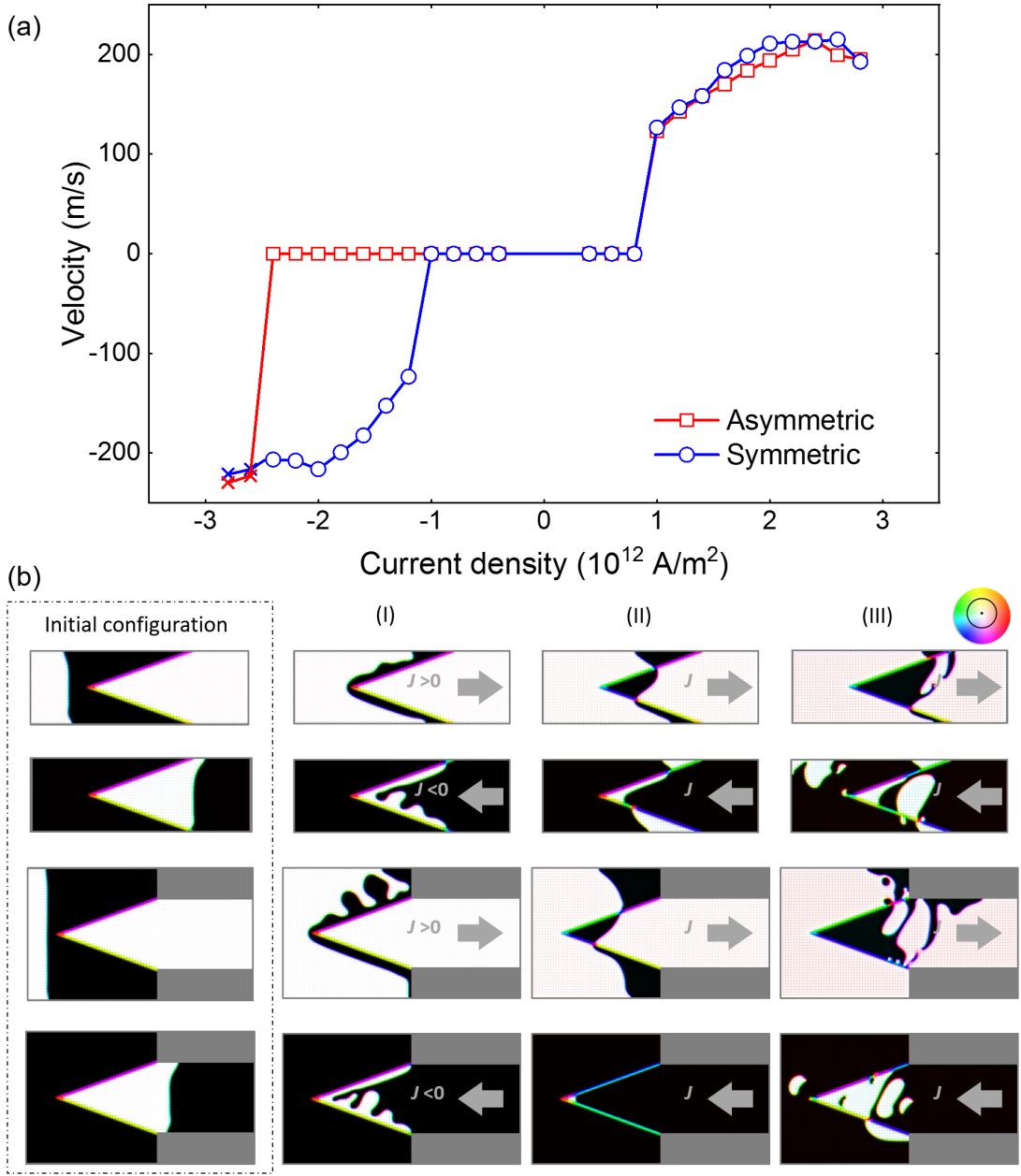

**Fig. 7.** Micromagnetic simulations of symmetric and asymmetric DW inverters. (a) Velocity of DWs transferring through the inverters as a function of current density. The crosses indicate the "breakdown" of the inversion process at high current density, corresponding to the continuous injection of ⊙|⊗ and ⊗|⊙ DWs. (b) Snapshots of the magnetic configuration of the symmetric (top two rows) and asymmetric inverter (bottom two rows) in regime (I) with $j= 4\times10^{11}$ A/m$^2$, regime (II) with $j= 1.6\times10^{12}$ A/m$^2$, and regime (III) with $j= 2.6\times10^{12}$ A/m$^2$. The direction of the magnetization is indicated by the color wheel.

**Figure 8**

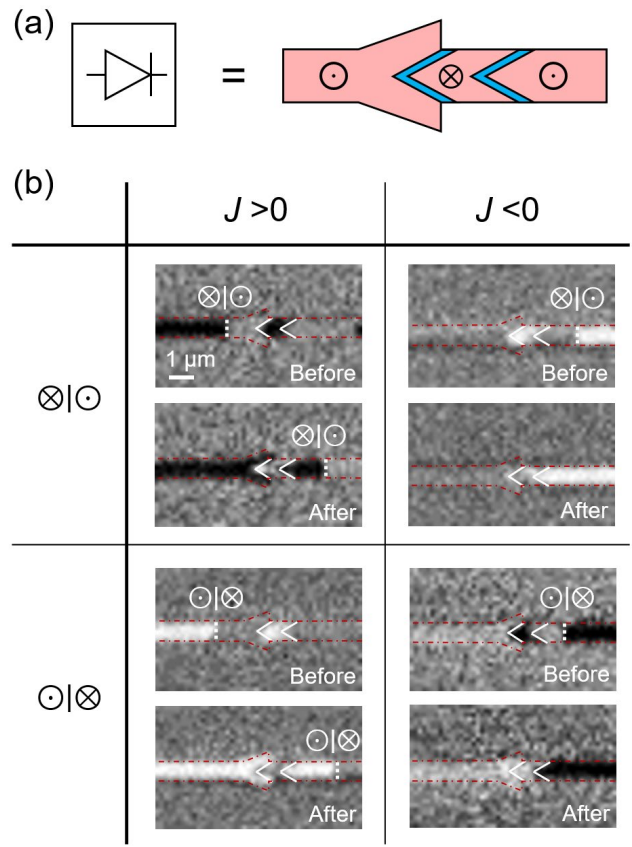

**Fig. 8.** (a) Symbol and schematic of the DW diode. (b) Differential MOKE images of the DW diode before and after applying 30 electric current pulses. The bright and dark contrast in the MOKE images corresponds to ⊙ and ⊗ magnetization, respectively. The current density and duration of the current pulses are $4.2\times10^{11}$ A/m$^2$ and 50 ns.